\documentclass[a4paper,12pt]{article}
\usepackage{amsmath}
\usepackage{amsfonts}
\usepackage{graphicx}

\begin{document}
\title{Quantum reversibility and \\a new model of quantum automaton}
\author{Massimo Pica Ciamarra}
\date{}
\maketitle
\begin{center}
\small{Dipartimento di Scienze Fisiche, Universit\`a
di Napoli ``Federico II'', Napoli, Italy. \\
E-mail: \textit{picaciam@na.infn.it}}
\end{center}

\begin{abstract}
This article is  an attempt to generalize the classical theory of
reversible computing, principally developed by Bennet [IBM J.
Res. Develop., 17(1973)] and by Fredkin and Toffoli [Internat. J.
Theoret. Phys., 21(1982)], to the quantum case. This is a
fundamental step towards the construction of a quantum computer
because a time efficient quantum computation is a reversible
physical process. The paper is organized as follows. The first
section reviews the classical theory of reversible computing. In
the second section it is showed that the designs used in the
classical framework to decrease the consumption of space cannot
be generalized to the quantum case; it is also suggested that
quantum computing is generally more demanding of space than
classical computing. In the last section a new model of fully
quantum and reversible automaton is proposed. The computational
power of this automaton is at least equal to that of classical
automata. Some conclusion are drawn in the last section.
\end{abstract}

\section{Reversible computing: a glance}
The classical theory of reversible computing has been analyzed
extensively (\cite{ben1, fretof}). The main results are the
following: 1. Every irreversible computation $f: x \rightarrow
f(x)$ can be effectively enclosed in a reversible computation $F:
(x,0) \rightarrow (f(x),x)$. $F$ is reversible because its input
$(x,0)$ is uniquely determined by its output $(f(x),x)$. 2. The
time and space required to compute $F$ are linearly dependent on
those required to compute $f$. These results imply that there are
no effects on the complexity hierarchy when the time and space
required to compute $f$ are those required to compute its
reversible realization $F$ (reversible space equal irreversible
space). To investigate whether is it possible to extend the above
classical results to the quantum case, we first revisit the ideas
behind classical reversible computing.

\subsection{Reversible gates}
In order to transform an irreversible gate into a reversible one,
not only the gate computation result is to be codified in the
gate output, but also some other information necessary to reverse
the computation (garbage). To store the garbage more output lines
are needed and, since a reversible gate has the same numbers of
input and output lines, even more input lines are required
(source). Note that, since reversibility implies that the output
of a reversible gate depends on the state of the source, its state
has to be properly set to compute a given boolean function by a
reversible gate. Figure 1 here below outlines the logical
structure of a reversible gate.
\begin{figure}[ht]
\begin{center}
\includegraphics[scale = 0.6]{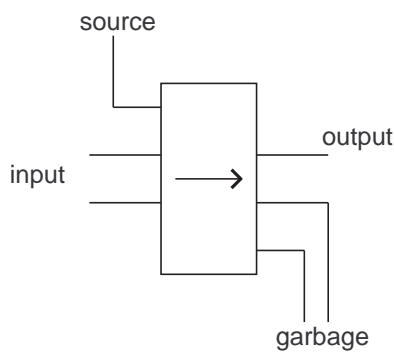}
\caption{Logical structure of a reversible gate}
\end{center}
\end{figure}

\noindent \textbf{Example: Reversible AND gate.} The AND gate is
a two-input one-output gate, whereas its reversible realization
is a three-input three-output gate. The adjunctive input line
($s$) constitutes the source and the two adjunctive output lines
($g_1$ and $g_2$) are the garbage. The transition function of the
reversible AND gate is reported in Table \ref{tab-revandgate}.
\begin{table}[h]
\caption{Reversible AND gate}
\[
\begin{array}{ccc|ccc}
s & i_1 & i_2 & AND & g_1 & g_2 \\
\hline 0 & 0 & 0 & 0 & 0 & 0 \\
0 & 0 & 1 & 0 & 0 & 1 \\
0 & 1 & 0 & 0 & 1 & 0 \\
0 & 1 & 1 & 1 & 1 & 1 \\
\end{array}
\]
\label{tab-revandgate}
\end{table}

\noindent Table \ref{tab-revandgate} could be completed by adding
the three possible inputs with $s=1$. In this case, however,
reversibility implies that the gate does not realize the AND
function. As already remarked, to use a reversible gate the value
of the source lines must be prepared appropriately.

\subsection{Reversible combinatorial circuits}
To build up a reversible combinatorial circuit (i.e. a circuit
without loop) from an irreversible one, every gate can be replaced
by its reversible realization (that is, every boolean function is
computable by a reversible circuit). However, although the final
circuit works, it often has too many source and garbage lines. In
fact \cite{fretof}:
\begin{quote}
\label{ciation-fretof}\textit{Note that, in general,} (in a
combinatorial circuit) \textit{the number of gates increases
exponentially with the number of input lines. This is so because
almost all Boolean functions are ``random", i.e., cannot be
realized by a circuit simpler than one containing an exhaustive
look-up table. Thus, in the ``wasteful" approach the amount of
garbage grows exponentially with the size of the argument.}
\end{quote}

\begin{figure}[h]
\begin{center}
\includegraphics[scale = 0.6]{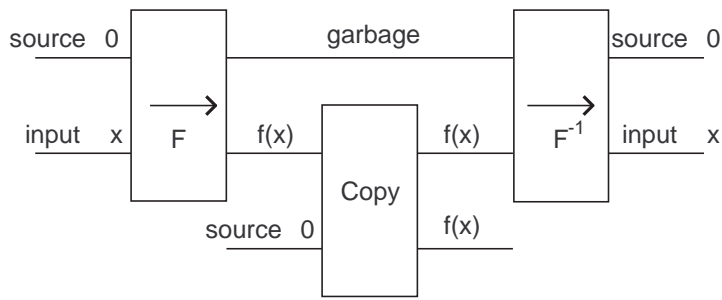}
\caption{Reversible combinatorial circuit with garbage recycle}
\label{fig-circseqrev}
\end{center}
\end{figure}

This explains why a reversible realization of a combinatorial
circuit seems less efficient than the combinatorial circuit
itself with regard to space consumption. Actually, in classical
case there is a way to recycle the garbage that implies a linear
relation between the space required to compute a boolean function
by a reversible or irreversible computation. This process,
outlined in Figure \ref{fig-circseqrev}, rests on two
considerations: 1. An inverse circuit exists for every reversible
circuit, and 2. A copy circuit: $(x,0) \rightarrow (x,x)$ does
exist.

\noindent To be honest, the above process is really useful only
when applied in a nested fashion.

\subsection{Reversible sequential circuits}
If the output lines of a combinatorial circuit are used as input
lines, one ends up with a sequential circuit (finite automaton),
as outlined in Figure 3. A sequential circuit constantly works in
(discrete) interaction with the environment, that changes the
state of the input lines. Its evolution is a three-step loop
starting from a given internal state:

\noindent 1. The environment sets the value of the input lines.
\noindent 2. The combinatorial circuit (i.e. the transition
function) is computed.
\noindent 3. Some output lines are used as
input lines (they codify the new internal state).

\begin{figure}[f]
\begin{center}
\includegraphics[scale = 0.8]{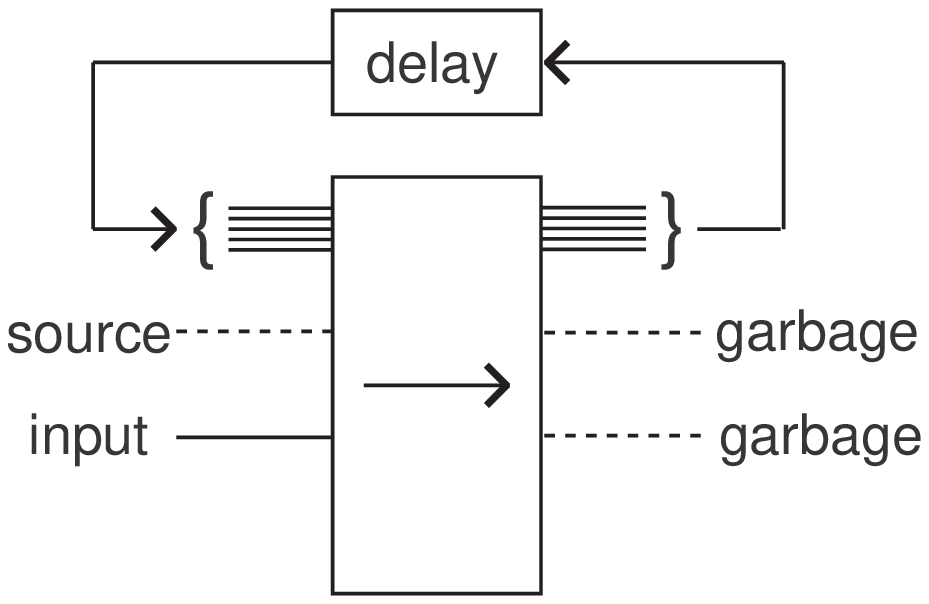}
\caption{Sequential circuit built on a reversible combinatorial
circuit.}
\end{center}
\end{figure}

By substituting the combinatorial part of a sequential circuit
for its reversible realization one does not end up with a
reversible sequential circuit. In fact, every time the
combinatorial circuit (transition function) is computed, the
garbage is modified and the information necessary to reverse the
computation is lost. Therefore, in order to build up a reversible
sequential circuit, every time the transition function is
computed the produced garbage has to be registered. In the
following section we present a scheme of sequential circuits, or
finite automata, that is able to register the output and the
garbage appropriately.

\subsection{Reversible Turing machines}
Bennet \cite{ben1} has proposed an effective procedure that
transforms every irreversible one-tape Turing machine into a
reversible three-tape Turing machine. The computation of this
machine, that recycles the garbage with the same trick used by
reversible combinatorial circuits (Fig. 2), proceeds as follows:

\begin{itemize}
\item[1.]Initially the three tapes contain \textit{input-source-source}.
The reversible function is computed.
\item[2.]The tapes contain \textit{garbage-result-source}. The $2^{nd}$ tape result is copied
on the $3^{rd}$ tape.
\item[3.]The tapes contain \textit{garbage-result-result}. The inverse reversible function
is computed.
\item[4.]The tapes contain \textit{input-source-result}.
\end{itemize}

\section{Space consumption of quantum computing}
\subsection{Space consumption of reversible computing}
In the classical framework we seldom care about space
consumption. In fact, since $P-Time \subseteq P-Space = NP-Space
\subseteq Exp-Time$, we first have to trouble about time
consumption. This relation stems from the fact that if a
computation needs some space, then it needs also some time to use
it. Therefore every consumption of space requires a consumption
of time, whereas the consumption of time does not implies a
consumption of space.

In the case of reversible computing, not only every consumption
of space requires a consumption of time, but also every
consumption of time requires a consumption of space. In fact every
unit of time, that corresponds to the execution of an elementary
step, as a (reversible) gate or a transition of a (reversible)
Turing machine, requires its own source. Time and space seem to
be strictly tied. Let $T$ and $S$ be the time and space required
by an irreversible Turing machine to end its computation,
supposing that the machine halts, and let $T'$ and $S'$ be the
time and space that its reversible realization does need. Then
exists such $c>1$ that $T' \simeq c+cT$ and $S' \simeq c+c(S+T)$
\cite{ben1}. From a computational complexity point of view this
suggests that reversible and irreversible computation differ. In
fact $S' = O(S+T)$: if an irreversible computation is space
efficient but time inefficient, then its reversible simulation is
both time and space inefficient.

However, as we saw in the previous section, Bennet has proposed a
way to reduce the amount of space the reversible simulation
requires. His scheme has been improved overtime \cite{Li}, and
today we know that reversible and irreversible computation have
the same complexity, i.e. $T' \simeq O(T)$ and $S' \simeq \simeq
S^2$). The consumption of space of a reversible computation does
not depend on the consumption of time of the irreversible one.
This result is appreciated because, although the need of space
$S' \simeq c +c(T+S)$ does not generally imply that the
reversible computation is space inefficient, nevertheless it
\textit{can be an unacceptable amount of space for many
practically computations} \cite{Li}.

\subsection{The quantum case}
The design proposed to recycle the garbage relies on the
possibility of retracing the computation, once duplicated the
result. In the quantum framework this design does not work
because of the no cloning theorem: it is not possible to make a
copy of the state of a quantum system. This is especially true
for a quantum computation that is more efficient than a classical
one, for it uses superposition states. On the contrary, if a
quantum computation is simply a small version of a classical
reversible computation, then this problem does not arise.

It could be possible to let quantum computation be less hungry of
space by relaxing reversibility. A quantum computation could
proceed by applying alternatively a reversible evolution operator
and an irreversible reuse of garbage. This way reversibility is
lost because the garbage is erased, but this is not warring
because we do not want to compute backward. With this scheme a
quantum system for information processing would act in a way
similar to that of the current model of quantum automata
\cite{Moore,Ambanis} (we will review it later on).

The problem of this design of garbage recycles stems from
entanglement. Because the output of a time efficient quantum
computation is entangled with the garbage, as in Shor's factoring
algorithm, the latter cannot be recycled without messing the
computation.

From these considerations we conclude that a quantum system that
computes a function $f$ consumes more space than a classical
system (reversible or not) that computes the same function.
Moreover, these considerations suggest that a quantum computation
is necessarily reversible, i.e., it must be possible to retrace
the input starting from the output (before the final
measurement), not only because of the unitarity of the evolution
operator, but also because output and garbage are generally
entangled.

\section{Quantum automata}
Since a quantum computational system is a generalization of a
classical one, it follows that its computational power should be
greater or equal than that of the classical system. However, the
computational power of quantum automata has been shown to be
smaller than that of classical automata. We argue that this
paradox origins from the fact that the currently accepted
definition of quantum automaton neglects the concept of quantum
reversibility. Here below, we revisit the role that reversibility
plays into quantum automata and propose a new model of quantum
finite automaton whose computational power is at least able to
recognize regular languages.

\subsection{The currently accepted definition of quantum automata}
The currently accepted definition of quantum finite automaton has
been proposed by Moore and Crutchfield \cite{Moore} and Ambanis
and Freivalds \cite{Ambanis}. A quantum finite automaton
\cite{Moore} $\mathcal{Q}$ consists of:
\begin{itemize}
\item[-] A Hilbert space $H$,
\item[-] An initial state vector $|s_{init}\rangle \in H$ with $\langle s_{init}|s_{init}\rangle=1$,
\item[-] A subspace $H_{accept} \subseteq H$ and an operator $P_{accept}$ that projects
into it,
\item[-] An input alphabet $A$, and
\item[-] A unitary transition matrix $U_a$ for each symbols $a \in A$.
\end{itemize}
Using the shorthand $U_w=U_{w_1}U_{w_2} \ldots U_{w_k}$ the
language accepted by $\mathcal{Q}$ is the function
$f^\mathcal{Q}(w)=|P_{accept}U_w |s_{init}\rangle|^2$ from words
in $A^*$ to probabilities in $[0,1]$.

The input alphabet consists of classical elements, whereas the
internal states are represented by quantum systems (by vectors of
a Hilbert space): this automaton is not fully quantum. Moreover,
it is neither reversible. In fact from the final state $U_w
|s_{init}\rangle$ one cannot trace back the computation because
$w$ is unknown. To trace back the computation one will need some
information that is not encoded in the final state. As to the
computational power of this kind of quantum automaton, first A.
Kondacs and J. Watrous \cite{Kondacs} have showed that there is a
regular language that cannot be accepted by a quantum finite
automaton. Then, A. Brodosky and N. Pippenger have demonstrated
that there is a whole class of regular languages that cannot be
accepted by quantum finite automata.

\subsection{(Quantum) Definite events}
\label{def-event} 
\begin{figure}[th]
\begin{center}
\begin{picture}(310,210)(0,0)
\put(10,0){\includegraphics[scale = 0.7]{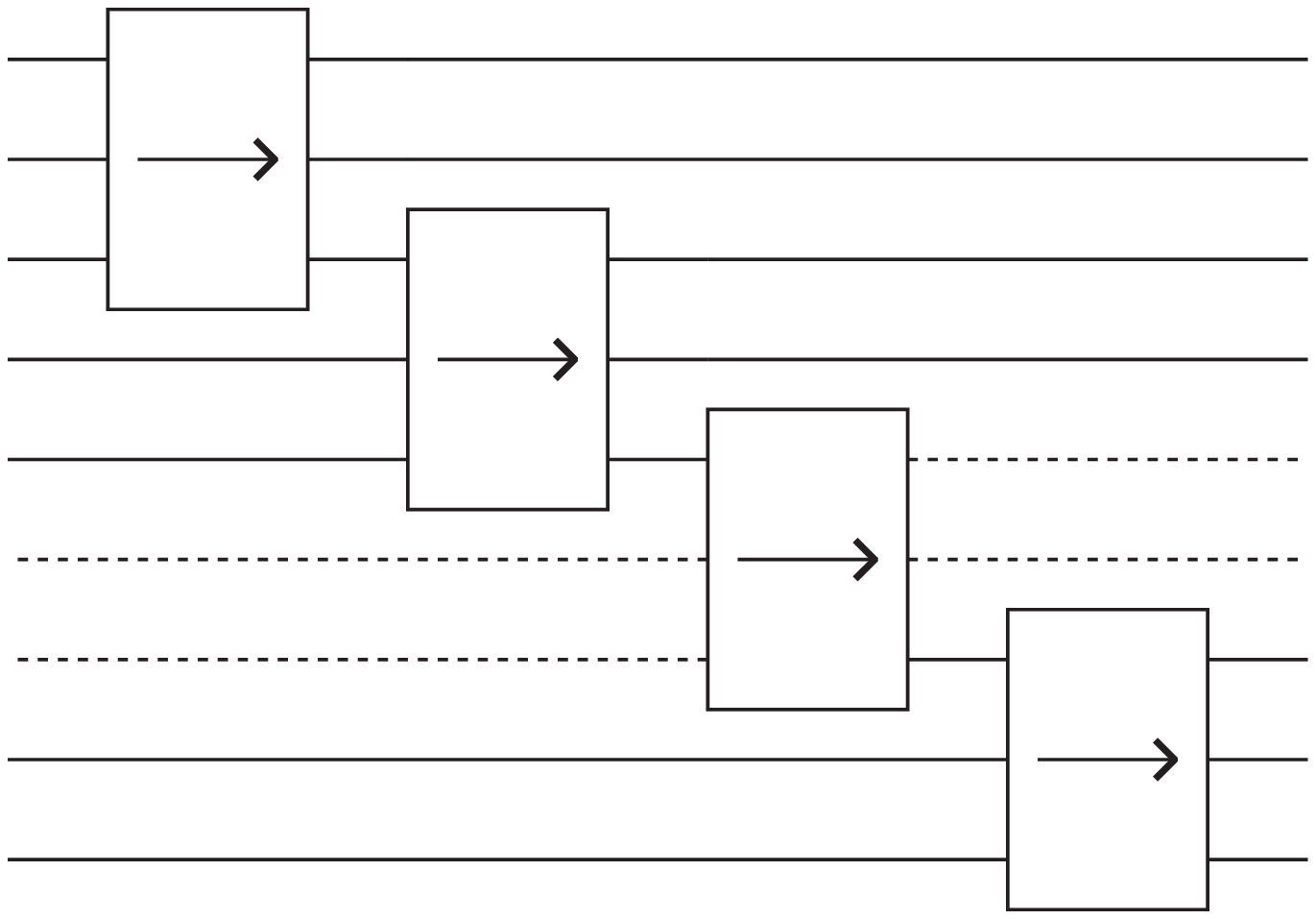}}
\put(246,45){$f$} \put(183,87){$f$} \put(120,130){$f$}
\put(57,171){$f$} \put(5,15){$s_n$} \put(5,36){$x_n$}
\put(5,57){$\ldots$} \put(5,78){$\ldots$} \put(5,99){$s_2$}
\put(5,121){$x_2$} \put(5,142){$s_1$} \put(5,163){$x_1$}
\put(5,184){$q$} \put(298,184){$g_1^1$} \put(298,163){$g_1^2$}
\put(298,142){$g_2^1$} \put(298,121){$g_2^2$}
\put(298,99){$g_3^1$} \put(298,78){$g_3^2$} \put(298,57){$\ldots$}
\put(298,36){$\ldots$} \put(298,15){$q_{end}$}
\end{picture}
\caption{Reversible combinatorial circuit that accepts definite
events.} \label{fig-defev}
\end{center}
\end{figure}

To build up a fully reversible model of quantum
automaton we begin setting up a reversible quantum system that
recognizes definite events\footnote{A definite event is a subset
of a regular event whose elements have fixed length.}. If
$\mathcal{L}_Q$ is the language accepted by the finite automaton
$\mathcal{Q}$, whose reversible transition function is $f$, then
the definite event $\mathcal{L}^n_Q$ is: $\mathcal{L}^n_Q=\{x | x
\in \mathcal{L}_Q \;and\; |x|= n\}$. $\mathcal{L}^n_Q$ is
recognized by the reversible combinatorial circuit of Figure
\ref{fig-defev}, where $q$ is the initial state, and $x_i$ and
$s_i$ are the $i^{th}$ input and his associated source
respectively. This circuit is set up by connecting many
combinatorial circuits that compute the transition function, each
with its own input and source. Of the three output lines of these
circuits, the first two ($g_i^1$ and $g_i^2$) contain the
garbage, while the third encodes the new internal state.
Reversibility is preserved by recording the value of the output
and garbage each time the transition function is computed.

We have assumed that every gate has three input lines. In general
you can suppose that every line represents a group of lines, so
as the internal states set and the input symbols set may contain
more that two elements.

Since the transition function is invertible this circuit could be
thought as a quantum circuit. In the classical case it is possible
to recycle the garbage either with the typical scheme, i.e.,
coping of the result and computing backward, or erasing the
garbage and using it as new source. These solutions, as we saw in
the previous section, are not generally useful in the quantum
case.

\subsection{A new model: an informal presentation}
To realize a quantum finite automaton we extend the previous
construction to the case of inputs of arbitrary finite length. We
do not want different systems, even if constructed uniformly, to
accept $\mathcal{L}_Q^1,\mathcal{L}_Q^2,\ldots,\mathcal{L}_Q^n$.
We want one system that accepts all of them simultaneously.

The solution consists in storing the input in a quantum tape, that
is a collection of equal quantum systems. Because of the presence
of the source, actually we introduce two quantum tapes: the input
and the source-garbage tape. This is exactly what Bennet did in
his model of reversible Turing machine. However we know that the
length of our tapes must be equal to that of the input sequence,
while in the Turing machine case the determination of the
maximums tape length is an undecidable problem.

Our automaton has a read/write head and an internal quantum
state. The head reads one cell on the input tape and a
corresponding cell on the source-garbage tape. Then, according to
the internal state and to the reversible transition function, it
changes the internal state and the states of the cells it read
(these cells become garbage). Then, it moves to the right and
read the next cell, and the loop restart. Our automaton acts like
a reversible two-tape Turing machine whose head is ever moving to
the right. It repeats its loop $n$ times, where $n$ is the length
of the input string.

\subsection{A new model: formal definition}
A quantum finite automaton $\mathcal{Q}$ consists of:
\begin{itemize}
\item[-] A Hilbert space $H_{in}$. This is the space of the internal states.
\item[-] An initial vector $|s_{0}\rangle \in H_{in}$ with $\langle s_0|s_0\rangle=1$.
\item[-] A subspace $H_{accept} \subseteq H_{in}$ and an operator $P_{accept}
$ that projects into it.
\item[-] A quantum tape $\mathcal{I}$ composed by a sequence of quantum systems whose
states are described in the Hilbert space $I$.
\item[-] A quantum tape $\mathcal{SG}$ composed by a sequence of quantum systems whose
states are described in the Hilbert space $SG$.
\item[-] A unitary evolution operator $U: H_{in}\otimes I \otimes SG\rightarrow H_{in} \otimes I\otimes SG$.
\end{itemize}

The input alphabet is composed by an orthonormal basis of $I$, a
quantum system whose dimension depend on the automaton we are
constructing. If we are generalizing to the quantum case a
classical reversible automaton $\mathcal{Q}$, then the dimension
of $I$ depends on the number of input symbols of this automaton.

The dimension of $SG$ is related to the way the unitary evolution
operator $U$ acts. In particular, if we want generalize to the
quantum case the reversible automaton $\mathcal{Q}$, constructed
from the irreversible automaton $\mathcal{Q}_{irr}$ following
Bennet's procedure, then the dimension of $SG$ is equal to the
number of elements on which the transition function of
$\mathcal{Q}_{irr}$ is defined (i.e., to the number of quadruples
of $\mathcal{Q}_{irr}$). In fact, to construct the reversible
version of $\mathcal{Q}_{irr}$, following Bennet \cite{ben1} we
have to introduce a new `garbage-symbol' for each transition of
$\mathcal{Q}_{irr}$.

The state of the $k^{th}$ `cell' of $\mathcal{I}$ is described by
a vector belonging to the space $I_k=I$, while the state of the
$k^{th}$ `cell' of $\mathcal{SG}$ is described by a vector
belonging to $SG_k=SG$. If the input string is $|w\rangle =
|w_1\rangle |w_2\rangle \ldots |w_n\rangle$, with $|w_k\rangle
\in I_k$, then the tape $\mathcal{SG}$ contains $|0_1\rangle
|0_2\rangle \ldots |0_n\rangle$, with $|0_k\rangle \in SG_k$.

\begin{figure}[bh]
\begin{center}
\begin{picture}(280,130)(0,0)
\put(0,0){\includegraphics[scale = 0.8]{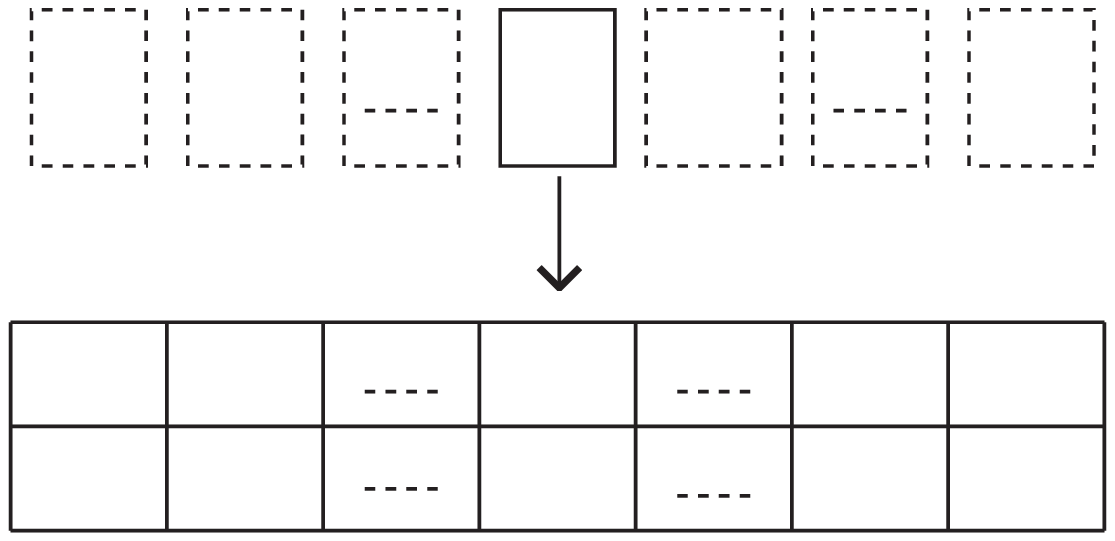}}
\put(18,15){$|b_1\rangle$} \put(18,40){$|a_1\rangle$}
\put(53,15){$|b_2\rangle$} \put(53,40){$|a_2\rangle$}
\put(123,15){$|b_i\rangle$} \put(123,40){$|a_i\rangle$}
\put(191,15){$|0_{n-1}\rangle$} \put(191,40){$|w_{n-1}\rangle$}
\put(231,15){$|0_n\rangle$} \put(231,40){$|w_n\rangle$}
\put(18,108){$|s_1\rangle$} \put(53,108){$|s_2\rangle$}
\put(125,108){$|s_i\rangle$} \put(234,108){$|s_n\rangle$}
\end{picture}
\caption{Quantum finite automata (particular case).}
\label{fig-rev-aut}
\end{center}
\end{figure}

In order to understand how the automaton works, we first suppose
that the states of the internal state, both $\mathcal{I}$ and
$\mathcal{SG}$ tapes, are continuously factorizing. In this case,
if $P_{in}$ projects into $H_{in}$, the internal state
$|s\rangle$ evolves as in Table \ref{tab-evolution}. The input
string $|w\rangle$ is accepted with probability
$\big{|}P_{accept}|s(n)\rangle\big{|}^2$.
\begin{table}[th]
\[
\begin{array}{c|l}
t & \hspace{3cm}|s(t)\rangle \\
\hline
1 & |s(1)\rangle = |s_0\rangle \\
2 & |s(2)\rangle = P_{in}U\big{(}|s(1)\rangle \otimes |w_1\rangle \otimes |0_1\rangle \big{)} \\
3 & |s(3)\rangle = P_{in}U\big{(}|s(2)\rangle \otimes |w_2\rangle \otimes |0_2\rangle \big{)} \\
\vdots & \hspace{3.5cm} \vdots \\
n & |s(n)\rangle = P_{in}U\big{(}|s(n-1)\rangle \otimes
w_{n-1}\rangle \otimes |0_{n-1}\rangle \big{)}
\end{array}
\]
\caption{Evolution of the quantum automaton internal state
(particular case).} \label{tab-evolution}
\end{table}
In this case the automaton is similar to a classical reversible
one\footnote{Note that in the literature a classical automaton
with a reversible evolution and an irreversible reuse of garbage
as new source (see Fig. 2) is frequently said `reversible' (as in
\cite{Toffoli1}). This is the definition of `reversible'
automaton that has been generalized to the quantum case
\cite{Ambanis, Moore}. We refer to another definition of
reversible classical automaton: a reversible Turing machine
\cite{ben1} whose head moves only to the right.}. Indeed, we can
schematize our automaton as in Figure \ref{fig-rev-aut}.

When the automaton is not anymore restricted the states of the
quantum register and those of the quantum tapes are not generally
factorizing. Although the mathematical description of the system
evolution is a little trickier, the automaton works in the same
way.

In this case it turns useful to define:
\[\widetilde{U_i}=\mathbb{I}_1 \otimes \mathbb{I}_2 \otimes \ldots \otimes \mathbb{I}_{i-1}
 \otimes \mathbb{I}_{i+1} \otimes \ldots \otimes \mathbb{I}_n \otimes U_i,\]
where $\mathbb{I}_k$ is the identity of $I_k \otimes SG_k$, $U_i$
the unitary operator $U$ when applied to $H_{in}\otimes I_i
\otimes SG_i$, and
\[|0\rangle^n = |0_1\rangle \otimes |0_2\rangle \otimes \ldots
\otimes |0_n\rangle, \;\; |0_k\rangle \in SG_k.\]

With this notation $|w\rangle = |w_1\rangle |w_2\rangle \ldots
|w_n\rangle$ is accepted with probability
\[P(|w\rangle)=\big{|}P_{accept}
\widetilde{U}_n \widetilde{U}_{n-1} \ldots \widetilde{U}_1
|s_0\rangle \otimes |w\rangle \otimes |0\rangle^n \big{|}^2.\]

It is worth noting that, although when you apply
$\widetilde{U}_k$ you formally act on the space
$\big{(}\bigotimes^n_{i=0} I_i \otimes SG_i \big{)} \otimes
H_{in}$, actually: 1. A local operation is performed, and 2. The
states of $I_i$ and $SG_i$ with $i>k$ do not modify. Therefore,
when $\widetilde{U}_k$ with $k<i$ is applied, you do not have to
dispose the `cells' described by $I_i$ or $SG_i$. This further
implies that at the beginning of the computation the tapes length
are not to be specified and that more `cells' can be added as the
computation moves on. This is exactly what happens in the
classical case.

\subsection{Computational power}
From \cite{ben1} we can deduce that every classical automaton can
be effectively transformed into a classical reversible automaton.
This reversible automaton acts on two tapes, and moves only to
the right. In fact, If we apply Bennet's procedure \cite{ben1} to
a Turing machine whose head move only to the right, we end up to
a reversible two-tape Turing machine that moves only to the right
(as already remarked in the quantum case we do not use the third
tape). Since our model of quantum automaton can clearly simulate
this classical reversible automaton, it follows that it has at
least the same computational power of classical automata.

\section{Conclusion and future developments}
In this paper we have revisited the classical theory of
reversible computing and studied its generalization to the
quantum case.

We have showed that it is not possible to generalize to the
quantum case the design that made classical reversible computation
space efficient (because of the `no cloning theorem'), and
pointed out that, because of the entanglement, it is not possible
to reuse the garbage without messing the computation. Moreover we
clarified that the quantum computing reversibility, namely the
possibility of tracing back the computation from the final
result, due to the unitarity of the quantum evolution, is not
eliminable because of the presence of non-local correlation.

These consideration had led us to propose a new model of fully
quantum and reversible automaton that has at least the same
computational power of classical automata. This automaton is,
however, `angry' of space. This is a relevant result since the
computational power of the currently accepted model of quantum
finite automata is thought to be smaller than that of classical
finite automata. The main difference between the quantum automata
here proposed and the currently accepted model of quantum
automata concerns reversibility.

In future the computational power of the proposed model of
quantum finite automaton should be investigated. Particularly, I
am interested in the role of the garbage. Dose the computational
power of the proposed model vary by allowing or preventing
interference between the final state and the garbage. In other
words: is the garbage useful, or is it simply a by-product of
quantum computing?

\subsection{Acknowledgments}
This work is for the most part based on my degree thesis. I
particularly tank my supervisor, Prof. Giuseppe Trautteur, and my
co-supervisor, Prof. Vincenzo R. Marigliano, both at the
Dipartimento di Scienze Fisiche dell'Universit\`a di Napoli
``Federico II''.
\\
\\
\noindent Universit\`a di Napoli ``Federico II'', Dipartimento di
Scienze Fisiche. Internal report $n^\circ \text{ DSF09/2001}$.
\small
\bibliographystyle{plain}
\bibliography{qc_bib}
\end{document}